
%
\input phyzzx.tex


\font\largemath = cmti10 scaled \magstep2

\textfont4=\largemath
\mathchardef\lS="0453

\footline={\hfill}

%








\def\overset#1\to#2{\mathop{#2}\limit^{#1}}
\def\underset#1\to#2{\mathop{#2}\limit_{#1}}
%
\REF\pw{G.Parisi and Y.Wu, Sci.Sin. {\bf 24} (1981) 483}
\REF\dh{For a review, P.H.Damgaard and H.H\"uffel, Phys.Rep. {\bf 152}
(1987) 227;\hfill\break
M.Namiki, {\it Stochastic Quantization} (Springer-Verlag, 1992)}
\REF\without{M.Namiki, I.Ohba, K.Okano and Y.Yamanaka, Prog. Theor. Phys.
{\bf 69} (1983) 1580;\hfill\break
  A.Nakamura, Prog. Theor. Phys. {\bf 86} (1991) 925}
\REF\nami{M.Namiki, I.Ohba and K.Okano, Prog. Theor. Phys. {\bf 72} (1984) 350}
\REF\ike{K.Ikegami, T.Kimura and R.Mochizuki, Nucl. Phys. {\bf B395} (1993)
371}
\REF\ps{S.Ryang, T.Saito and K.Shigemoto, Prog. Theor. Phys. {\bf 73}
(1985) 1295;\hfill\break
A.M.Horowitz, Phys. Lett. {\bf 156B} (1985) 89}
\REF\ohba{I.Ohba, Prog. Theor. Phys.  {\bf 77} (1987) 1267}
\REF\ito{K.Ito, Proc. Imp. Acad. {\bf 20} (1944) 519}
\REF\gra{R.Graham, Phys. Lett. {\bf 109A} (1985) 209;\hfill\break
H.Rumpf, Phys. Rev. {\bf D33} (1986) 942}
\REF\ly{T.D.Lee and C.N.Yang, Phys. Rev. {\bf 128} (1962) 885}
\REF\mochi{R.Mochizuki, Prog. Theor. Phys. {\bf 85} (1991) 407}
\REF\moch{R.Mochizuki, Prog. Theor. Phys. {\bf 88} (1992) 1233}
\REF\dirac{P.A.M.Dirac, {\it Lectures on Quantum Mechanics} (Yeshiva
Univ.,1964); \hfill\break
L.D.Faddeev, Theor. Math. Phys. {\bf 1} (1970) 1}

%
%
\unnumberedchapters
\date={}
%
\pubnum={CHIBA-EP-71}
\date={October 1993}
\titlepage
\title{The Stochastic Quantization Method in Phase Space and a New Gauge
Fixing Prodedure}

\author{{\it Riuji} Mochizuki\footnote*{e-mail address:
mochi@cuphd.nd.chiba-u.ac.jp}}
\address{Department of Physics, Chiba
University, \break
          1-33 Yayoi-cho, Inage-ku, Chiba 263, Japan}

\baselineskip = 12pt
\footline={\hfill-- \folio\ --\hfill}
\baselineskip = 18pt
\section{Abstract}

We study the stochastic quantization of the system with first class
constraints in phase space.  Though the  Langevin equations of
the canonical variables are defined without ordinary gauge fixing procedure,
gauge fixing conditions are automatically selected and introduced by imposing
stochastic consistency conditions upon the first class constraints.  Then the
equilibrium solution of the Fokker-Planck equation is identical with
corresponding path integral distribution.  \vfill\eject

\section{$\S$1. INTRODUCTION}

The stochastic quantization method, which has been proposed by Parisi and
Wu\refmark\pw in 1981, is known as the third quantization method\refmark\dh
following the canonical and  path integral formalism.  In the scheme
fields $\phi(x)$ have an additional coordinate $t$;
$$
\phi(x)\rightarrow\phi(x,t),
$$
coordinate which is called the fictitious time.  The development of the
fields with the lapse of the fictitious time  is described by a Langevin
equation $$
\eqalignno{
d\phi(x,t)
& \equiv\phi(x,t+dt)-\phi(x,t) \cr
& =-{\partial {\rm L}\over\partial \phi}\mid_{\phi=\phi(x,t)}+dW(x,t), }
$$
where L is the Lagrangian of a system and $dW$ is the Wiener process whose
 correlations  are defined as $$
\langle dW(x,t)\rangle =0,
$$
$$
\langle dW(x,t)dW(x^{\prime},t)\rangle =2\delta(x-x^{\prime})dt.
$$
The equations enable us to calculate  correlation functions of the
fields, which tend to  corresponding Green functions in the equilibrium
limit $t\rightarrow\infty$.

The correlation functions may be given in another form
$$
\langle\cdots\rangle=\int D\phi\cdots P(\phi ,t),
$$
where $P$ is a probability distribution. A Fokker-Planck equation, which
$P$ satisfies, is obtained by calculating, with the Langevin equation, the
correlation function of  the fictitious time derivative of any functional
$f(\phi)$ and comparing it  with\break $\int D\phi f\cdot\partial P/\partial
t$.  The Fokker-Planck equation thus becomes
$$
{\partial P\over\partial t}={\partial\over\partial\phi}\Bigl\{{\partial
{\rm L}\over\partial \phi}+{\partial\over\partial\phi}\Bigr\} P.
$$
The equation has in the equilibrium limit $\partial P/\partial
t=0$ a solution
$$
P=\exp (-{\rm L}),
$$
which is the same as corresponding path integral distribution.

It is one of the large merits of the stochastic quantization method that it
enables us to quantize gauge fields without gauge fixing
conditions.\refmark\without  Hence if we extend the stochastic quantization
method, which is ordinary defined in configuration space, to phase space, we
may shed light on the quantization of the systems with first class constraints
(FCC).  Until now Namiki et al.\refmark\nami and we\refmark\ike
have studied how to handle constraints by the stochastic quantization method
in configuration space (the SQM-cs), showing that it is much useful to impose
the consistency condition with the lapse of the fictitious time, the
stochastic consistency condition, upon the constraints.  On the other hand,
some proposals for the stochastic quantization method in phase space (the
SQM-ps) have been made.\refmark\ps   Ohba\refmark\ohba has tried to quantize
by the SQM-ps, in particular,
 the system with FCC where  gauge fixing conditions are added.
It is accordingly the system  with second class constraints that he
quantized by  the SQM-ps.

In this paper we quantize the system with FCC by the SQM-ps without  usual
gauge fixing procedure and show that  gauge fixing conditions
 are automatically introduced through the stochastic consistency
conditions upon the FCC.  Since the prescription to impose them is
straightforward and has little dependence on the form of the FCC, it will
 be useful when we quantize complicatedly constrained fields and may bring us
the other knowledge about gauge theory than the other quantization methods
have offered.

 This paper is organized as follows.  In $\S$2, taking notice of
the relation between the SQM-cs and the SQM-ps and, in particular, the Lee-Yang
term, we define the strict form of the Langevin equations with a constant
kernel, of  canonical variables.  In $\S$3, we show how to apply the SQM-ps
to the system with FCC.  We impose the stochastic consistency conditions upon
the FCC and  the determinant factor which appears with the gauge fixing
procedure consequently becomes included in the solution of the Fokker-Planck
equation in the equilibrium
limit.  In $\S$4, we define new fields and
divide them into constraint variables  and independent ones to clarify what
gauge the stochastic consistency conditions accord to.
 $\S$5 is devoted to the conclusion.   Besides them, we give
in the appendix how to apply our formulation to the system with second class
constraints (FCC and gauge fixing conditions).

\section{$\S$2. THE LANGEVIN EQUATION IN PHASE SPACE}

In this section we consider the system with the Lagrangian in  Euclidean
space-time
$$
{\rm L}={1\over 2}G_{AB}(q)\partial_{\mu}q^A\partial_{\mu}q^B,\eqno(2.1)
$$
$$
A, B=1,\cdots ,N,
$$
$$
\mu=0,\cdots ,n.
$$
For such a system as includes a field-dependent metric, by using Ito's
calculation rule,\refmark\ito the Langevin equation in configuration space
is
$$
dq^A(t)=-G^{AB}{\partial {\rm L}\over\partial q^B}dt+{1\over
\sqrt{G}}{\partial\over\partial q^B}(\sqrt{G}G^{AB})dt+E^A_{\
a}dW^a,\eqno(2.2)
$$
where $G^{AB}$ is the inverse of the metric $G_{AB}$ and $E^A_{\ a}$ is the
vielbein defined with its inverse $E_A^{\ a}$ as
$$
E^A_{\ a}E^B_{\ b}\delta^{ab}=G^{AB},
$$
$$
E_A^{\ a}E_B^{\ b}\delta_{ab}=G_{AB},
$$
$$
E^A_{\ a}E_B^{\ a}=\delta^A_B,\ \ \ \ \ \ \ E^A_{\ a}E_A^{\ b}=\delta^b_a,
$$
$$
a, b=1,\cdots ,N.: {\rm reference\ frame\  indices}
$$
The additional term of the Langevin equation (2.2) is necessary for general
coordinate transformation (GCT) covariance\refmark\gra and, in perturbation,
cancellation of divergent terms.\refmark\ike  Moreover we can observe it
induces the Lee-Yang term\refmark\ly in
 equilibrium Fokker-Planck distribution.\refmark\mochi  The Fokker-Planck
equation in this case is
$$
{\partial P\over \partial t}={\partial\over\partial
q^A}G^{AB}\Bigl\{{\partial{\rm L}\over\partial
q^B}-{1\over\sqrt{G}}{\partial\sqrt{G}\over\partial
q^B}+{\partial\over\partial q^B}\Bigr\}P,
$$
$$
G\equiv\det(G_{AB}),
$$
the equilibrium solution of which is
$$
P=\exp(-{\rm L}+{1\over 2}lnG).
$$
The correlation function is thereby written as
$$
\langle\cdots\rangle=\int Dq\cdots \exp(-{\rm L}+{1\over 2}lnG),
$$
which is consistent with the path integral formalism.

Let us shift to phase space.  First we define the conjugate momentum $p_A$ of
$q^A$ and the Hamiltonian $H$ as
$$
p_A=+i{\partial {\rm L}\over \partial (\partial_0
q^A)}=iG_{AB}\partial_0q^B,\eqno(2.3)
$$
$$
H=+ip_A\partial_0 q^A+{\rm L}={1\over 2}G^{AB}p_Ap_B+{1\over
2}G_{AB}\partial_iq^A\partial_iq^B.\eqno(2.4)
$$
The equations of motion in phase space are the Hamilton equations
$$
-i\partial_0 p_A-{\partial H\over \partial q^A}=0,\eqno(2.5a)
$$
$$
+i\partial_0 q^A-{\partial H\over \partial p_A}=0,\eqno(2.5b)
$$
in place of the Euler-Lagrange equation in configuration space.  It is
natural to adopt the LHS of the Hamilton equations (2.5) as naive drift terms
of the Langevin equations in the  phase space.  Moreover since the metric of
the manifold spanned by the canonical variables $q$ and $p$ is $$
\left(\matrix{
\hfill G_{AB}(q) & \hfil 0\hfill\cr
\hfill 0 & \hfil G^{AB}(q) \hfill\cr}\right),
$$
determinant of which is unity,  the GCT covariant Langevin equations are
defined in our formulation as
$$
\left\{\eqalign{
dq^A&=G^{AB}(-i\partial_0p_B-{\partial H\over\partial
q^B})dt+{\partial G^{AB}\over\partial q^B}dt+E^A_{\ a}dW^a,\cr
dp_A&=CG_{AB}(+i\partial_0q^B-{\partial H\over\partial
p_B})dt+\sqrt{C}E_A^{\ a}dV_a,
\cr
}\right.\eqno(2.6)
$$
where $C$ is the constant kernel determined later and  $dW^a$ and $dV_a$ are
the Wiener process defined to have following correlations:
$$
\langle
dW^a(x,t)dW^b(x^{\prime},t)\rangle=2\delta^{ab}\delta^n(x-x^{\prime})dt,
$$
$$
\langle dV_a(x,t)dV_b(x^{\prime},t)\rangle
=2\delta_{ab}\delta^n(x-x^{\prime})dt,
$$
$$
\langle dW^a\rangle=\langle dV_a\rangle=\langle dW^adV_b\rangle=0.
$$
The probability distribution $P(q,p,t)$ in the phase space is also defined
through correlation functions as
$$
\langle \cdots\rangle=\int DqDp\cdots P(q,p,t).
$$
Similar calculation to the case in the configuration space leads the\break
Fokker-Planck equation
$$
{\partial P\over\partial t}=\Bigl[{\partial\over\partial
q^A}G^{AB}\Bigl\{{\partial L\over\partial q^B}+{\partial\over\partial
q^B}\Bigr\}+CG_{AB}{\partial\over\partial p_A}\Bigl\{{\partial
L\over\partial p_B}+{\partial\over\partial p_B}\Bigr\}\Bigr]P,\eqno(2.7)
$$
where $L$ is the Lagrangian in the phase space defined as
$$
L\equiv -ip_A\partial_0q^A+H.
$$
The Fokker-Planck equation has a solution in the equilibrium limit with which
the correlation function is rewritten as
$$
\langle\cdots\rangle=\int DqDp\cdots\exp(-L).
$$

The constant kernel $C$,  introduced for
dimensional consistency between the Langevin equations (2.6), has the same
dimension as $t^{-1}$.    It is fixed if we demand that the Langevin
equations in the phase space (2.6) are reduced to the  Langevin equation in
the configuration space (2.2).  For the purpose,  we put the Hamiltonian (2.4)
into Eqs.(2.6):
$$
\left\{\eqalign{
dq^A&=\Bigl[-iG^{AB}\partial_0p_B-{G^{AB}\over 2}\big({\partial
G^{CD}\over\partial q^B}p_Cp_D+{\partial G_{CD}\over\partial
q^B}\partial_iq^C\partial_iq^D\big)\cr
&\ \ +G^{AB}\partial_i(G_{BC}\partial_iq^C)+{\partial G^{AB}\over\partial
q^B}\Bigr]dt+E^A_{\ a}dW^a,\cr
dp_A&=C\Bigl[iG_{AB}\partial_0q^B-p_A\Bigr]dt+\sqrt{C}E_A^{\
a}dV_a.\cr}\right.\eqno(2.8)
$$
We eliminate the momenta in Eqs.(2.8) and obtain the Langevin equation of
$q^A$
$$
\eqalignno{dq^A=\Bigl[
&-iG^{AB}\partial_0(iG_{BC}\partial_0q^C+{E_B^{\
a}dV_a\over\sqrt{C}dt}-{dp_B\over Cdt})-{G^{AB}\over 2}\big\{{\partial
G_{CD}\over\partial q^B}\partial_iq^C\partial_iq^D\cr
&+{\partial G^{CD}\over\partial
q^B}(iG_{CE}\partial_0q^E+{E_C^{\
a}dV_a\over\sqrt{C}dt}-{dp_C\over
 Cdt})(iG_{DF}\partial_0q^F+{E_D^{\
b}dV_b\over\sqrt{C}dt}-{dp_D\over Cdt})\big\}\cr
&+G^{AB}\partial_i(G_{BC}\partial_iq^C)+{\partial G^{AB}\over\partial
q^B}\Bigr]dt+E^A_{\ a}dW^a.&(2.9) }
$$
Taking the correlation of the Wiener process $dV_a$ in the limit
$dt\rightarrow 0$, Eq.(2.9) should be identical with (2.2).  It is realized if
$$
C={2\over dt},\eqno(2.10)
$$
which is also supported in view of the dimension.  We therefore conclude that
the Langevin equations of the canonical variables $q^A$ and $p_A$ are
$$
\left\{\eqalign{
dq^A&=-G^{AB}{\partial L\over\partial
q^B}dt+{\partial G^{AB}\over\partial q^B}dt+E^A_{\ a}dW^a,\cr
dp_A&=-2G_{AB}{\partial L\over\partial
p_B}+\sqrt{{2\over dt}}E_A^{\ a}dV_a.
\cr
}\right.\eqno(2.11)
$$
The drift term (the first term on the RHS) of the latter equation of (2.11)
being very powerful because of the order of $dt$, $p_A$ will reaches an
equilibrium state in a very short time.  It justifies the manipulation to
connect Eq.(2.2) with Eq.(2.9); taking the correlation only of $dV_a$ accords
to thinking the system at a finite time.    The discussion on the constant
kernel in this section may also be applied when we write the Langevin equation
of an auxiliary  field, the kinetic term of which does not exist in the
Lagrangian.

\section{$\S$3.  FIRST CLASS CONSTRAINTS AND STOCHASTIC CONSISTENCY
CONDITIONS}

In this section we consider the system, the Lagrangian of which is
L$(q^I,\partial_0q^I)$ $(I=1,\cdots  ,N)$, with FCC
$$
F^a(q,p)=0,  \ \ \ \ \ \ \ \ \ (a=1,\cdots ,M).\eqno(3.1)
$$
The conjugate momentum $p_I$ and the Hamiltonian $H$ are defined through Eqs.
(2.3) and (2.4) respectively and the metric of the system is assumed to be
$\delta_{IJ}$ for simplicity.  Following the canonical approach the Hamiltonian
is generalized to include the FCC as
$$
H_G\equiv H+\lambda_aF^a,\eqno(3.2)
$$
where $\lambda_a$ is a multiplier field, which cannot be determined without
gauge fixing in the canonical formalism.  We define the Langevin
equations accordingly as
$$
\eqalign{
dq^I&=-{\partial L_G\over\partial q^I}dt+dW^I,\cr
dp^I&=-C{\partial L_G\over\partial p^I}dt+\sqrt{C}dV^I,\cr}
$$
where the singular Lagrangian $L_G$ is
$$
L_G\equiv -ip^I\partial_0q^I+H_G.\eqno(3.3)
$$
It is convenient to define symplectic variables
$$
\phi^i\equiv\{ {p^1\over\sqrt{C}},\cdots ,{p^N\over\sqrt{C}},q^1,\cdots
,q^N\},
$$
$$
dY^i\equiv\{dV^1,\cdots ,dV^N,dW^1,\cdots ,dW^N\},
$$
with which we can rewrite the Langevin equation in a simple form
$$
d\phi^i=-{\partial L_G\over\partial \phi^i}dt+dY^i.\eqno(3.4)
$$
When we quantize gauge field with the Langevin equation in configuration space,
we do not need to fix the gauge freedom;  the Faddeev-Popov ghost effects are
automatically carried into correlation functions and divergence, though
appearing in the correlation functions, does not  in physical
quantities.\refmark\without  The Langevin equation (3.4), however,  is not
suitable for perturbation because of the multipliers, and neither is
 the Fokker-Planck formulation.  The Fokker-Planck equation derived through
the Langevin equation (3.4) is
$$
{\partial P\over\partial
t}={\partial \over\partial \phi^i}\Bigl\{{\partial
L_G\over\partial\phi^i}+{\partial \over\partial\phi^i}\Bigr\} P,\eqno(3.5)
$$
whose solution in the equilibrium limit is
$$
P=\exp (-L_G).\eqno(3.6)
$$
We cannot obtain proper correlation functions either by using the distribution
(3.6).

We need to change the Langevin equation (3.4) into the form which fits for
our handling.  We, following the case in the configuration
space,\refmark\nami\refmark\ike try to impose on each FCC (3.1) the stochastic
consistency condition $$
dF^a=0\eqno(3.7)
$$
and hence determine the multiplier $\lambda_a$.  Our strategy may seem
strange since the multipliers for FCC cannot be fixed in the canonical
formalism.  We will however accomplish the purpose as the following.
If we  take Ito's formula\refmark\ito into consideration, the stochastic
consistency condition (3.7) becomes
$$
\eqalignno{
0&=dF^a\cr
&={\partial F^a\over\partial \phi^i}d\phi^i+{1\over 2}{\partial^2
F^a\over\partial\phi^i\partial\phi^j}\langle d\phi^id\phi^j\rangle_{O(dt)}\cr
&={\partial F^a\over\partial \phi^i}\Bigl\{\Bigl( -{\partial
L\over\partial\phi^i}-\lambda_b{\partial F^b\over\partial\phi^i}\Bigr)
dt+dY^i\Bigr\}\cr &\ \ \ \ +{1\over 2}{\partial^2
F^a\over\partial\phi^i\partial\phi^j}\langle
d\phi^id\phi^j\rangle_{O(dt)},&(3.8)}
$$
where $\langle
d\phi^id\phi^j\rangle_{O(dt)}$ is the correlation whose order is  $dt$.  It
is not $2\delta^{ij}dt$ since $\lambda_a$ contains the Wiener process $dY^i$
and hence after setting the multipliers concretely we will compute
it.\refmark\ike\refmark\moch  We change Eq. (3.8) into
$$
D^{ab}\lambda_b= -{\partial F^a\over\partial\phi^i}{\partial
L\over\partial\phi^i}+{1\over 2dt}{\partial^2 F^a\over\partial\phi^i\partial
\phi^j}\langle d\phi^id\phi^j\rangle_{O(dt)}+{\partial
F^a\over\partial\phi^i}{dY^i \over dt},\eqno(3.9)
$$
where
$$
D^{ab}\equiv{\partial F^a\over\partial\phi^i}{\partial
F^b\over\partial\phi^i}.\eqno(3.10)
$$
Assuming that  $D^{ab}$ has the inverse $D^{-1}_{ab}$, we obtain the definite
form of $\lambda_a$
$$
\lambda_a=D^{-1}_{ab}
\Bigl\{ -{\partial F^b\over\partial\phi^i}{\partial
L\over\partial\phi^i}+{1\over 2dt}{\partial^2 F^b\over\partial\phi^i\partial
\phi^j}\langle d\phi^id\phi^j\rangle_{O(dt)}+{\partial
F^b\over\partial\phi^i}{dY^i \over dt}\Bigr\}.\eqno(3.11)
$$
The Langevin equaiton (3.4), into which Eq.(3.11) is put, becomes
$$
d\phi^i=K^{ij}\Bigl\{ -{\partial L\over\partial\phi^j}dt+dY^j\Bigr\}
-{1\over 2}{\partial F^a\over\partial\phi^i}D^{-1}_{ab}{\partial^2
F^b\over\partial \phi^j\partial\phi^k}\langle
d\phi^jd\phi^k\rangle_{O(dt)},\eqno(3.12)
$$
where $K^{ij}$ is the projection operator which extracts the quotient space
composed of only the independent variables:
$$
K^{ij}\equiv\delta^{ij}-R^{ij},
$$
$$
R^{ij}\equiv{\partial F^a\over\partial\phi^i}D^{-1}_{ab}{\partial
F^b\over\partial\phi^j}.
$$
In this stage we can compute $\langle
d\phi^id\phi^j\rangle_{O(dt)}$ by using the Langevin equation (3.12) and
finally obtain the Langevin equation without the multipliers:
$$
d\phi^i=K^{ij}\Bigl\{ -{\partial  L\over\partial\phi^j}dt+dY^j\Bigr\}
-{\partial F^a\over\partial\phi^i}D^{-1}_{ab}{\partial^2
F^b\over\partial \phi^j\partial\phi^k}K^{jk}dt,\eqno(3.13)
$$
which is the desirable form for calculation and of course satisfies Eq. (3.7).
Imposing the stochastic consistency condition thereby enables us to
determine the multipliers for the FCC, which fact suggests that the stochastic
consistency condition may be equivalent to some gauge fixing condition.

 The Fokker-Planck equation, expected to bring us more information, is
$$
{\partial P(\phi, t)\over\partial
t}={\partial\over\partial\phi^i}K^{ij}\Bigl\{ {\partial
L\over\partial\phi^j}-{\partial^2
F^a\over\partial\phi^j\partial\phi^k}D^{-1}_{ab}{\partial
F^b\over\partial\phi^k}+{\partial\over\partial\phi^j}\Bigr\} P(\phi,
t)\eqno(3.14)
$$
Taking the FCC (3.1) into consideration, we  write the equilibrium
solution as
$$
P=\int D\lambda\sqrt{D}\exp(-L-\lambda_aF^a),\eqno(3.15)
$$
$$
D\equiv\det(D^{ab}),
$$
where appears the determinant factor, which does not exist in the previous
solution (3.6).  We hence expect that the FCC (3.1) and the gauge fixing
conditions, which we will write as $C^a=0$,   due to the stochastic
consistency conditions (3.7), constitute the second class constraints whose
Poisson brackets satisfy
$$
{\rm det}\big\{ F^a,C^b\big\}_{P.B.}=\sqrt{D}.\eqno(3.16)
$$
The probability distribution (3.15), however, seems to include no gauge fixing
term and diverge in the same manner as the previous one (3.6).  Is the guess
wrong ?  If not,  we must find the finite solution of the Fokker-Planck
equation (3.14).  The answer will be obtained in the next section.

\section{$\S$4.  THE STOCHASTIC CONSISTENCY CONDITION AS GAUGE FIXING}

Our purpose in this section is to confirm the above mentioned conjecture by
investigating in more detail the Langevin equation (3.13) and the Fokker-Planck
equation (3.14), for which reason it is necessary to introduce new symplectic
variables and divide them into constraint and independent variables.  We write
the new variables as
$$
\Phi^{\mu}\equiv\{{\Pi^1\over\sqrt{C}},\cdots,{\Pi^N\over\sqrt{C}},Q^1,
\cdots,Q^N\},\ \ \ \ \ \ \ \ (\mu=1,\cdots,2N)\eqno(4.1)
$$
where $Q$ and $\Pi$ are canonical conjugate of each other.  Vielbein fields
are, connecting the new variables with the old ones, defined as
$$
E^{\mu}_{\ i}\equiv{\partial\Phi^{\mu}\over\partial\phi^i},\eqno(4.2)
$$
$$
E^{\mu}_{\ i}E_{\nu ,i}=\delta^{\mu}_{\nu},\ \ \ \ E^{\mu}_{\ i}E_{\mu
,j}=\delta_{ij}.\eqno(4.3)
$$
Besides the definition, since $Q$ and $\Pi$ are the canonical variables, the
vielbein fields must  satisfy the following relations:
$$
\left\{\eqalign{
E^{\alpha +N}_{\ \ \ \ \ I+N}&=E_{\alpha ,I},\cr
E^{\alpha +N}_{\ \ \ \ \ I}\ \  \ &=-E_{\alpha ,I+N},\cr
E_{\alpha +N,I+N}&=E^{\alpha}_{\ \ I},\cr
E_{\alpha +N,I}\ \ \ &=-E^{\alpha}_{\ \ I+N}.\cr
}\right.\eqno(\alpha=1,\cdots ,N;I=1,\cdots ,N)\ \ \ \
(4.4)
$$
Each of the constraints $F^a$, assumed to be the first
class, may be transfered\refmark\dirac to the new momentum
$\Pi^a/\sqrt{C}$, that is, we set the vielbein fields (4.2) with the left-hand
indices $\mu=a\  (a=1,\cdots,M)$ as
$$
E^a_{\ i}={\partial F^a\over\partial\phi^i},\ \ \ \
E_{a,i}=D^{-1}_{ab}{\partial F^b\over\partial\phi^i}.\eqno(4.5)
$$
{}From Eqs. (4.4) and (4.5) we set those with $\mu=a+N$:
$$
\left\{\eqalign{
E^{a+N}_{\ \ \ \ \ I}\ \ \ &=-D^{-1}_{ab}{\partial
F^b\over\partial\phi^{I+N}},\cr
E^{a+N}_{\ \  \ \ \ I+N}&=D^{-1}_{ab}{\partial F^b\over\partial\phi^I},\cr
E_{a+N,I}\ \ \ &=-{\partial F^a\over\partial\phi^{I+N}},\cr
E_{a+N,I+N}&={\partial F^a\over\partial\phi^I}.\cr
}\right.\eqno(4.6)
$$
There are some more relations which Eqs. (4.4) and (4.5) yield:
$$
\left\{\eqalign{
E^a_{\ i}E^b_{\ i}\ \ &=E_{a+N,i}E_{b+N,i}=D^{ab},\cr
E_{a,i}E_{b,i}\ \ &=E^{a+N}_{\ \ \ \ i}E^{b+N}_{\ \ \ \ i}=D^{-1}_{ab},\cr
E^a_{\ i}E^{b+N}_{\ \ \ \  i}&=E_{a,i}E^{b+N}_{\ \ \ \ i}=E^a_{\
i}E_{b+N,i}=E_{a,i}E_{b+N,i}=0,\cr
R_{ij}&=E^a_{\ i}E_{a,j}
,\cr
K_{ij}&=E^{a+N}_{\ \ \ \ i}E_{a+N,j}+E^A_{\ i}E_{A,j}+E^{A+N}_{\ \ \ \
i}E_{A+N,j},\cr
&(A=M+1,\cdots,N).\cr
}\right.\eqno(4.7)
$$
The third equation of (4.7) is obtained with the help of the assumption that
the constraints (3.1) are the first class.  Moreover  $E^A_{\ i}$
and $E^{A+N}_{\ \ \ \ i}$ chosen orthogonal to each other and to the other
vielbein fields, the metric of the manifold spanned by the independent
variables $\Phi^A,\Phi^{a+N}$ and $\Phi^{A+N}$ is
$$
G_{\mu^{\prime}\nu^{\prime}}=\left(
\matrix{
\hfill g_{AB}&\hfil 0\hfil &0\hfill\cr
\hfill 0&\hfil D^{ab}\hfil &0\hfill\cr
\hfill 0&\hfil 0 \hfil &g^{AB}\hfill\cr
}\right),\eqno (\mu^{\prime}\nu^{\prime}=M+1,\cdots,2N)\ \ \ \ \ \ \ \ \ (4.8)
$$
with
$$
g^{AB}=E^A_{\ i}E^B_{\ i}=E_{A+N,i}E_{B+N,i},
$$
$$
g_{AB}=E_{A,i}E_{B,i}=E^{A+N}_{\ \ \ \ \ i}E^{B+N}_{\ \ \ \ \ i}.
$$
We thus obtain the Langevin equations of the new independent variables:
$$
\eqalignno{
d\Phi^{\mu^{\prime}}&={\partial
\Phi^{\mu^{\prime}}\over\partial\phi^i}d\phi^i+{\partial^2\Phi^{\mu^{\prime}}
\over\partial\phi
^i\partial\phi^j}K^{ij}dt\cr
&=G^{\mu^{\prime}\nu^{\prime}}\Bigl\{ -{\partial  L\over\partial
\Phi^{\nu^{\prime}}}dt+E_{\nu^{\prime},j}dY^j\Bigr\} +{\partial
E^{\mu^{\prime}}_{\ i}\over\partial\phi^j}K^{ij}dt,&(4.9)}
$$
while the constraint variables of course satisfy $d\Phi^a=0$.  Here we
assume\refmark\ike
$$
\nabla_{\mu^{\prime}}E^{\mu^{\prime}}_{\
i}\equiv{1\over\sqrt{D}}{\partial\over\partial\Phi^{\mu^{\prime}}}
(\sqrt{D}E^{\mu^{\prime}}_{\ i})=0,
\eqno(4.10)
$$
where we use $\det (G_{\mu^{\prime}\nu^{\prime}})=D$ and
$\nabla_{\mu^{\prime}}$ is the covariant derivative in the manifold spanned by
the independent variables, which assumption naturally leads  metric condition
$\nabla_{\mu^{\prime}}G^{\mu^{\prime}\nu^{\prime}}=0$.  The Langevin equation
(4.9) with Eqs. (4.8) and (4.10) is thereby reduced to
$$
d\Phi^{\mu^{\prime}}=G^{\mu^{\prime}\nu^{\prime}}\Bigl\{ -{\partial
L\over\partial \Phi^{\nu^{\prime}}}dt+E_{\nu^{\prime},j}dY^j\Bigr\}
+{1\over\sqrt{D}}{\partial\over\partial\Phi^{\nu^{\prime}}}(\sqrt{D}G^{\mu^{
\prime}\nu^{\prime}})dt,\eqno(4.11)
$$
which is GCT covariant in the manifold.

The Fokker-Planck
equation made up of the independent variables  is
$$
{\partial P\over\partial
t}=\Bigl[{\partial\over\partial\Phi^{\mu^{\prime}}}G^{\mu^{\prime}\nu^{\prime}}\Bigl\{
{\partial {\rm L}\over\partial\Phi^{\nu^{\prime}}}-{1\over\sqrt{D}}{\partial
\sqrt{D}\over\partial\Phi^{\nu^{\prime}}}+{\partial\over\partial\Phi^{\nu^
{\prime}}}\Bigr\}\Bigr] P,\eqno(4.12)
$$
which is no more than the rewriting of Eq. (3.14) and has only the same
solution.  Calculating the expectation value only of physical quantities, we
can however change the Fokker-Planck equation (4.12) into the desirable form
which has a finite solution.  Physical quantity  must not depend on
the value of the multipliers $\lambda_a$, that is, the Poisson bracket of FCC
and the physical quantity $\tilde f$ should be zero;
$$
\big\{\tilde f,F^a\big\}_{P.B.}={\partial \tilde
f\over\partial\Phi^{b+N}}=0,\eqno(4.13)
$$
which yields
$$
\langle d\tilde f\rangle=\langle {\partial \tilde
f\over\partial\Phi^A}d\Phi^A +{\partial\tilde
f\over\partial\Phi^{A+N}}d\Phi^{A+N}+{\partial^2\tilde
f\over\partial\Phi^A\Phi^B}g^{AB}+{\partial^2\tilde
f\over\partial\Phi^{A+N}\Phi^{B+N}}g_{AB}\rangle.
$$
The Fokker-Planck equation (4.12) hence loses the {\it gauge} variables
$\Phi^{a+N}$ and is reduced to
$$
\eqalignno{
{\partial \tilde P\over\partial
t}&=\Biggl[{\partial\over\partial\Phi^A}g^{AB}\Bigl\{  {\partial
L\over\partial\Phi^B}-{1\over\sqrt{D}}{\partial\sqrt{D}\over\partial\Phi^B}
+{\partial\over\partial\Phi^B}\Bigr\}\cr
&\ \ +
{\partial\over\partial\Phi^{A+N}}g_{AB}\Bigl\{ {\partial
L\over\partial\Phi^{B+N}}-{1\over\sqrt{D}}{\partial\sqrt{D}\over
\partial\Phi^{B+N}}
+{\partial\over\partial\Phi^{B+N}}\Bigr\}\Biggr]\tilde P.&(4.14)
}
$$

Finding the finite solution identical with corresponding
path integral distribution,  we decompose the vielbein fields as
$$
\left\{
\eqalign{
E^{\mu^{\prime}}_{\ i}&\equiv e^{\mu^{\prime}}_{\
\tilde\mu}\epsilon^{\tilde\mu}_{\ i},\cr
E_{\mu^{\prime},i}&\equiv e_{\mu^{\prime}}^{\
\tilde\mu}\epsilon_{\tilde\mu,i},\cr
}\right.\eqno(\tilde\mu =M+1,\cdots,2N)\ \ \ \ \ \ (4.15)
$$
with
$$
\epsilon^{\tilde\mu}_{\ i}\epsilon^{\tilde\nu}_{\ i}
=\delta^{\tilde\mu\tilde\nu},
$$
$$
\epsilon^{\tilde\mu}_{\
i}\epsilon_{\tilde\nu,i}=\delta^{\tilde\mu}_{\tilde\nu},\
$$
$$
\epsilon_{\tilde\mu,i}\epsilon_{\tilde\nu,i}=\delta_{\tilde\mu\tilde\nu},
$$
$$
e^A_{\ \tilde a+N}=e^{A+N}_{\ \ \tilde a+N}=e^{ a+N}_{\ \ \tilde
A}=e^{a+N}_{\ \ \tilde A+N}=0.
$$
We then define new variables $C^{\tilde a+N}\
(\tilde a=1,\cdots,M)$ as
$$
{\partial C^{\tilde a+N}\over\partial \Phi^{a+N}}\equiv e_{a+N}^{\ \ \tilde
a+N},\eqno(4.16)
$$
with which we can write the equilibrium solution of
the Fokker-Planck equation (4.14), setting $C^{\tilde a+N}=0$, as
$$
\tilde P=\int D\lambda
D\tilde\lambda\sqrt{D}\exp(-L+\lambda_aF^a+\tilde\lambda_{\tilde
a+N}C^{\tilde a+N}).\eqno(4.17)
$$
Taking notice that
$$
\det\{C^{\tilde a+N},F^a\}_{P.B.}=\det(e_{a+N}^{\ \ \tilde
a+N})=\sqrt D,
$$
the Fokker-Planck distribution (4.17) is identical with the   path
integral distribution which we obtain by imposing the gauge fixing conditions
$$
C^{\tilde a+N}=0,\ \ \ \ \ (\tilde a=1,\cdots,M)
$$
upon the FCC
$$
F^a=0.
$$
We thereby insist  that our conjecture in $\S$3 has been
confirmed.

\section{$\S$5. CONCLUSION }

In this paper we have formulated the SQM-ps on the classical basis of the
canonical formalism; the Langevin equations in the phase space reduce to
Hamilton equations in the classical limit.  Furthermore our prescription of
introducing constraints into the Langevin equations is consistent with Dirac's
formalism.  The Langevin equations with FCC, however, do not need to contain
ordinary gauge fixing conditions, in place of which stochastic consistency
conditions have been imposed and  enabled to determine the Lagrange
multipliers for the FCC.  It has suggested that the stochastic consistency
conditions should correspond to some gauge fixing.  More information has
been brought through the equilibrium  Fokker-Planck distribution:  After
imposing the stochastic consistency conditions upon the FCC, we can obtain the
finite values of physical quantities though the other quantities may
diverge in the limit $t\rightarrow\infty$.   We therefore conclude that
imposing stochastic consistency conditions on   FCC, being a little differnt
from
ordinary gauge fixing procedure, makes the FCC
change into second class constraints.

\section{ACKNOWLEDGEMENTS}
I would like to  thank  K. Ikegami for stimulating
discussions.

\section{ APPENDIX}

Here we show that our formulation is applicable to the systems with second
class constraints.  Following our prescription with gauge-fixed Hamiltonian
$$
\eqalignno{
\tilde H_G&=H-\lambda_aF^a-\tilde\lambda_{\tilde a+N}C^{\tilde a+N}
\cr
&\equiv H-\overline\lambda_{\rm a}\overline F^{\rm a},\cr
&({\rm a}=1,\cdots,2M)}
$$
we can straightforwardly reach the Langevin equation without the multipliers:
$$
d\phi^i=\overline K^{ij}\Bigl\{ -{\partial  L\over\partial\phi^j}dt+dY^j\Bigr\}
-{\partial \overline F^{\rm a}\over\partial\phi^i}\overline
D^{-1}_{{\rm ab}}{\partial^2 \overline F^{\rm b}\over\partial
\phi^j\partial\phi^k}\overline K^{jk}dt,
$$
where
$$
\overline D^{{\rm ab}}\equiv{\partial \overline F^{\rm a}\over\partial\phi^i}
{\partial\overline F^{\rm b}\over\partial\phi^i},
$$
$$
\overline K^{ij}\equiv\delta^{ij}-{\partial\overline
F^{\rm a}\over\partial\phi^i}\overline D^{-1}_{\rm
ab}{\partial\overline F^{\rm b}\over\partial\phi^j}.
$$
We can easily set up the Fokker-Planck equation, too:
$$
{\partial \overline P(\phi, t)\over\partial
t}={\partial\over\partial\phi^i}\overline K^{ij}\Bigl\{ {\partial
L\over\partial\phi^j}-{\partial^2
\overline F^{\rm
a}\over\partial\phi^j\partial\phi^k}\overline D^{-1}_{\rm ab}{\partial
\overline F^{\rm b}\over\partial\phi^k}+{\partial\over\partial\phi^j}\Bigr\}
\overline P(\phi, t),
$$
which has an equilibrium solution
$$
\overline P=\int D\overline\lambda\sqrt{\overline D}\exp(-L+\overline
\lambda_{\rm a}\overline F^{\rm a}).\eqno(A.1)
$$
The solution (A.1) should be identical with corresponding path integral
distribution, which we can confirm with the following equations:
$$
\eqalignno{
\det\{ C^{\tilde a+N},F^b\}_{P.B.}&=\det \Bigl({\partial C^{\tilde
a+N}\over\partial \Phi^{b+N}}\Bigr)\cr
&=\det\Bigl( e_{b+N}^{\ \ \tilde a+N}\Bigr),}
$$
$$
\eqalignno{ \det\overline D^{\rm ab}
&=\det D^{ab}\cdot\det\Bigl(\epsilon^{\tilde a+N}_{\ \ i}\epsilon^{\tilde
b+N}_{\ \ i}\Bigr)\cr
&=\det\Bigl( e_{b+N}^{\ \ \tilde a+N}\Bigr)^2.}
$$

\vfill\eject
\refout\vfill\eject
\end